\documentclass{pazhb_eng}
\usepackage[hyperfootnotes=false]{hyperref}
\usepackage{color}
\usepackage{epsf}
\usepackage[utf8]{inputenc}
\usepackage[T2A]{fontenc}
\usepackage{amsmath}
\usepackage{amsfonts}
\usepackage{amssymb}
\usepackage{graphicx}
\usepackage{gensymb}
\usepackage{float}
\usepackage{longtable}
\usepackage{changes}
\usepackage{wrapfig}
\usepackage{comment}
\usepackage{threeparttable}
\newcommand{\obj}{V379~Vir}

\begin{document}

\journalinfo{2025}{51}{2}{1}[0]
\UDK{///}

\title{\bf On accretion in the polar V379~Vir}
\author{
    M.V.~Suslikov\email{mvsuslikov@outlook.com}\address{1,2}, 
    A.I.~Kolbin\email{mvsuslikov@outlook.com (M.V. Suslikov), kolbinalexander@mail.ru (A.I. Kolbin)}\address{1},
    N.V.~Borisov\address{1}
    \addresstext{1}{Special Astrophysical Observatory of the Russian Academy of Sciences, Nizhny Arkhyz, Karachay-Cherkessia, Russia}
    \addresstext{2}{Kazan (Volga Region) Federal University, Kremlyovskaya St. 18, Kazan, Russia}
}


\shortauthor{M.V.~Suslikov et al.}

\shorttitle{On accretion in the polar V379~Vir}

\begin{abstract}
Based on optical and infrared survey data spanning $\approx 20$~years of observations, the long-term variability of the polar {\obj} with a brown dwarf secondary has been studied. By modeling the spectral energy distribution, we constrain the white dwarf’s mass to $M_1 = 0.61 \pm 0.05~M_{\odot}$ and its effective temperature to $T_{\mathrm{eff}} = 10930 \pm 350~\mathrm{K}$. Near-infrared photometry yields a donor radius of $R_2 = 0.095 \pm 0.018~R_{\odot}$ and temperature $T_{\mathrm{eff}} = 1600 \pm 180~\mathrm{K}$. Modeling of the cyclotron emission from the accretion spot, detected with the \textit{Spitzer} infrared telescope, gives an accretion rate of $\dot{M} \approx 3 \times 10^{-13}~M_{\odot}/\mathrm{yr}$. This rate is consistent with polars in a low accretion state, but significantly higher than expected from wind-driven mass transfer.

\englishkeywords{Stars: cataclysmic variables, polars (AM Her-type), brown dwarfs – Individual: V379~Vir (SDSS J121209.31+013627.7) – Methods: photometry.}

\end{abstract}

\section*{Introduction}
\label{Introduction}
\noindent
Cataclysmic variables are semi-detached binaries consisting of an accreting white dwarf (the primary) and a red or brown dwarf donor (the secondary), which fills its Roche lobe \citep{Warner95, Hellier01}. Mass transfer occurs through the inner Lagrange point L$_1$. In systems without a strong magnetic field ($B \lesssim 0.1$~MG), the transferred material forms an accretion disk around the white dwarf. The orbital evolution of cataclysmic variables is driven by angular momentum losses via magnetic braking (through a magnetized stellar wind from the donor) and gravitational wave emission \citep{Knigge11}. As the system evolves, the orbital period decreases until it reaches a minimum $P_{\mathrm{orb}} \approx 80$~min. At this point, the donor becomes degenerate, and further mass loss causes it to expand, leading to an increase in the orbital period. Such systems are known as \textit{period bouncers}.

A subclass of cataclysmic variables featuring highly magnetized white dwarfs ($B\sim 10-100$~MG) are polars (AM Her-type stars). In these binaries, the magnetic field suppresses the formation of an accretion disk, directing the ionized gas flow toward the magnetic poles of the accretor \citep{Cropper90}. The strong magnetic field also synchronizes the rotation of the white dwarf with the orbital motion. Typical accretion rates in polars are $\dot{M} \sim 10^{-11} - 10^{-9}$~M$_{\odot}/\mathrm{yr}$. Under such conditions, a shock front forms at the white dwarf's surface, heating the gas to $\sim 10-50$~keV. The gas then cools and settles, producing an accretion spot. Cooling proceeds primarily via two mechanisms: thermal bremsstrahlung in X-rays and cyclotron emission in the optical and infrared bands.

Within AM Her-type stars, there is a subclass of low accretion rate polars (LARP). These systems persist for extended periods in a low accretion state $\dot{M} \sim 10^{-14} - 10^{-13}$~M$_{\odot}/\mathrm{yr}$, corresponding to local accretion rates of $\dot m \sim 0.001-0.01~\mathrm{g}~\mathrm{cm^{-2}}~\mathrm{s^{-1}}$. In this regime, the infalling material does not form a shock; instead, it directly heats the atmosphere of the white dwarf to temperatures of $\sim 1$~keV. Cooling occurs primarily via cyclotron radiation \citep{Kuijpers82, Woelk92}. A distinguishing feature of LARPs is the presence of strong, well-separated cyclotron harmonics in optical and infrared spectra \citep[e.g.,][]{Schmidt05a, Roestel24}, along with very low X-ray luminosities $L_X \lesssim 10^{29}~\mathrm{erg/s}$. Some LARPs are thought to be pre-cataclysmic binaries with magnetic white dwarfs, gradually evolving into classical polars as they lose angular momentum. These are often referred to as pre-polars. In such systems, accretion is believed to occur via the donor’s stellar wind, funneled onto the poles by the magnetic field --- a process known as magnetic siphoning \citep{Schmidt05b}.

However, LARP-like observational features are also seen in systems that are not pre-polars. Often, these are typical AM Her-type stars that have temporarily entered a low accretion state. These are interpreted as episodes of reduced mass transfer, possibly due to localized magnetic activity of the donor. Intriguingly, LARP features also appear in period bouncers. Although evolutionary models predict that $40-70\%$ of cataclysmic variables should be period bouncers \citep{Belloni18}, observations suggest their fraction is closer to $\sim 10$\% \citep{Pala20, Inight23}. Recently, \citet{Schreiber23} proposed that in systems with very low accretion rates and $P_{\mathrm{orb}} \approx 80$~min, the crystallization of the white dwarf’s core can enhance its magnetic field. Interaction between the magnetic fields of the white dwarf and the donor can increase orbital angular momentum, leading to detachment. This hypothesis motivates further searches for LARPs among period bouncers, which may represent such detached systems.

The polar {\obj} (also known as SDSS~J121209.31+013627.7) is a confirmed period bouncer with $P_{\mathrm{orb}} = 88.4$~min, and has been studied extensively. \citet{Schmidt05a} first identified the system as a cataclysmic variable with a magnetic white dwarf ($B \approx 7$~MG) and a brown dwarf of spectral type L. To explain the low accretion rate, they proposed the system is detached and originated from a main-sequence star of mass $\sim 1.5$~M$_{\odot}$ and a brown dwarf orbiting at $\sim 1$~AU ($P_{\mathrm{orb}} \sim 1~\mathrm{yr}$). Follow-up infrared observations confirmed a cool donor ($T_\mathrm{eff} \lesssim 1700$~K) and variable cyclotron emission \citep{Debes06, Farihi08}. \textit{XMM-Newton} observations revealed variability modulated by the white dwarf’s rotation, consistent with the presence of an accretion spot \citep{Stelzer17}. From the X-ray flux, an accretion rate of $\dot{M} \sim 10^{-14}M_\odot/\mathrm{yr}$ was inferred; however, this estimate neglects the cyclotron component of the accretion luminosity. Furthermore, the white dwarf mass in the system remains poorly constrained. 

In this paper, we present an analysis of archival ultraviolet, optical, and infrared observations to further investigate the nature of {\obj}. First, we examine the long-term variability of {\obj} to search for possible accretion state transitions. Second, we model the optical and ultraviolet spectral energy distribution (SED) to refine the parameters of the white dwarf. Third, we estimate the donor's radius and Roche-lobe filling factor using near-infrared photometry. Finally, we determine the accretion rate in {\obj} based on infrared SED modeling of the cyclotron component.

\section{Long-term photometry}

\begin{figure*}[h!]
    \centering 
    \vspace*{3mm}
    \includegraphics[width=1.0\textwidth]{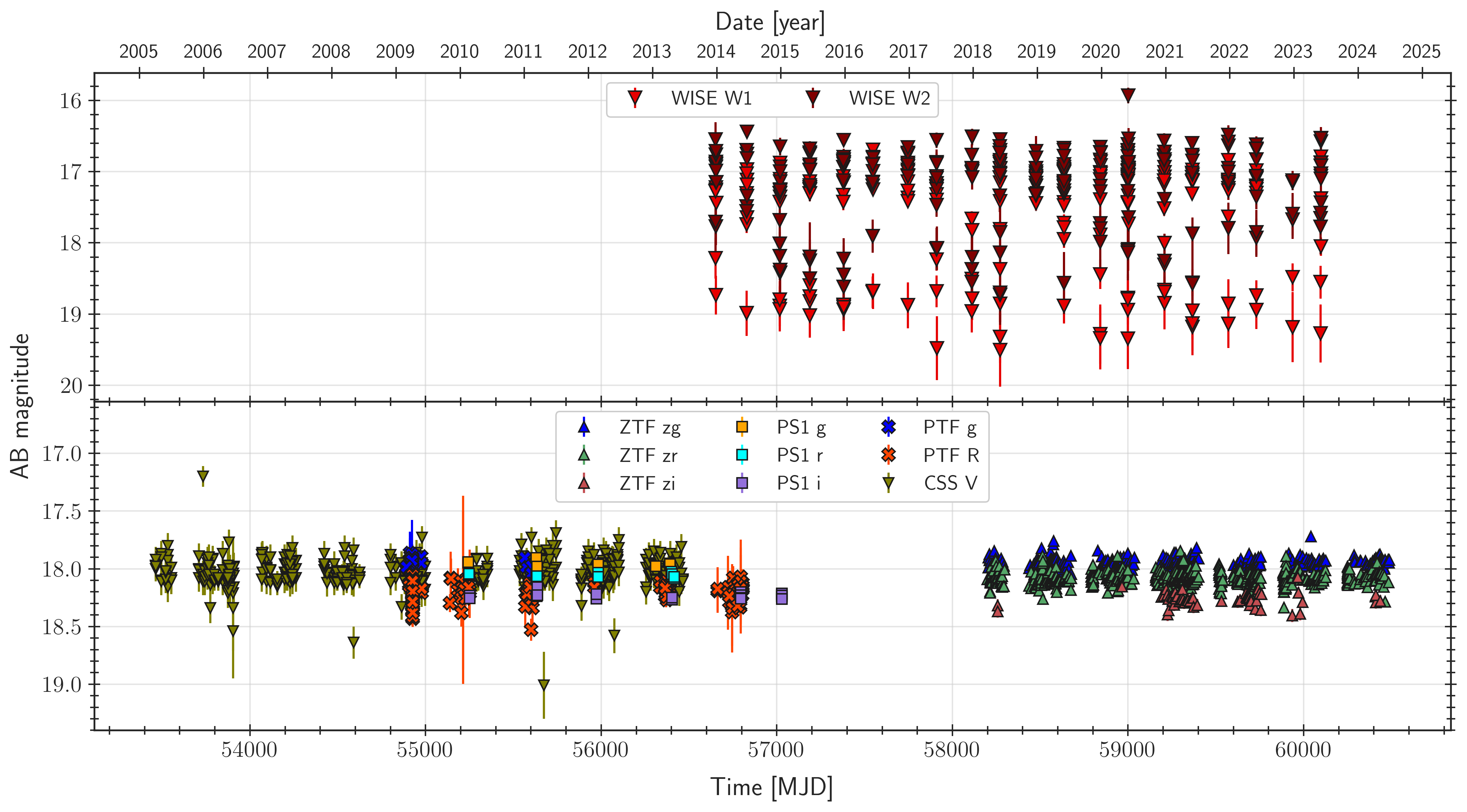} 
    \vspace*{-6mm}
    \caption{Long-term optical (bottom panel) and infrared (top panel) light curves of {\obj}. Photometric data from Pan-STARRS ($g$, $r$, $i$), ZTF ($g$, $r$, $i$), PTF ($R$), CSS ($V$) and WISE ($W1$, $W2$) are shown.}
    \label{fig:longterm_lc}
\end{figure*}

We investigated the long-term brightness variability of {\obj} using data from wide-field sky surveys, including the Catalina Sky Survey (CSS; \citealt{Drake09}), the Zwicky Transient Facility (ZTF; \citealt{Masci18}), the Palomar Transient Factory (PTF; \citealt{Law09}), and the Panoramic Survey Telescope and Rapid Response System (Pan-STARRS; \citealt{Flewelling20}). These datasets cover nearly two decades of observations and are compiled into a composite light curve shown in Fig.~\ref{fig:longterm_lc}. Over this timescale, {\obj} shows no significant optical variability above $\gtrsim 0.5~\mathrm{mag}$, which would otherwise indicate transitions between accretion states typical for AM~Her-type systems. The absence of state changes is further supported by data from the NEOWISE photometric catalog \citep{Mainzer14}, which provides mid-infrared observations from the \textit{WISE} telescope spanning $\approx 9.4$~years. Light curves in the $W1$ ($\lambda_{\rm eff} \approx 3.3~\mu$m) and $W2$ ($\lambda_{\rm eff} \approx 4.6~\mu$m) bands, shown in Fig.~\ref{fig:longterm_lc}, are dominated by cyclotron emission from the accretion spot. Since the intensity of cyclotron emission is directly linked to the accretion rate, the mid-IR observations clearly indicate no significant change in mass transfer over time. The apparent variability amplitude ranging from $\approx 16.5$~mag to $\approx 19.5$~mag arises from orbital modulation, as confirmed by phase-folded light curves in Fig.~\ref{fig:phased_lc}. Thus, {\obj} exhibits a common property of LARPs -- the absence of significant variability over long timescales \citep{Webbink05}.

To refine the orbital period of the system, we analyzed the $W2$-band light curve, which provides both long coverage and relatively high variability amplitude. Using a Lomb–Scargle periodogram \citep{VanderPlas18}, we derive an updated orbital period of $P_{\rm orb} = 0.0614070(7)$~days (or $88.426\pm0.001$~min), in agreement with earlier estimates by \citet{Burleigh06} and \citet{Debes06}. The revised ephemeris for {\obj} is:
\begin{gather}\label{ephemeris}
    \mathrm{BJD} = 2453798.6036(5) + 0.0614070(7) \times E,
\end{gather}
where the initial epoch corresponds to the midpoint of the bright phase in the light curves. We also note that a periodogram analysis of ZTF data reveals no evidence for non-radial pulsations in the white dwarf, as previously suggested by \citet{Schmidt05b}.

\section{Spectral energy distribution}

\begin{figure}[h!]
    \centering
    \begin{minipage}[t]{\columnwidth}
    \centering
    \includegraphics[width=\linewidth]{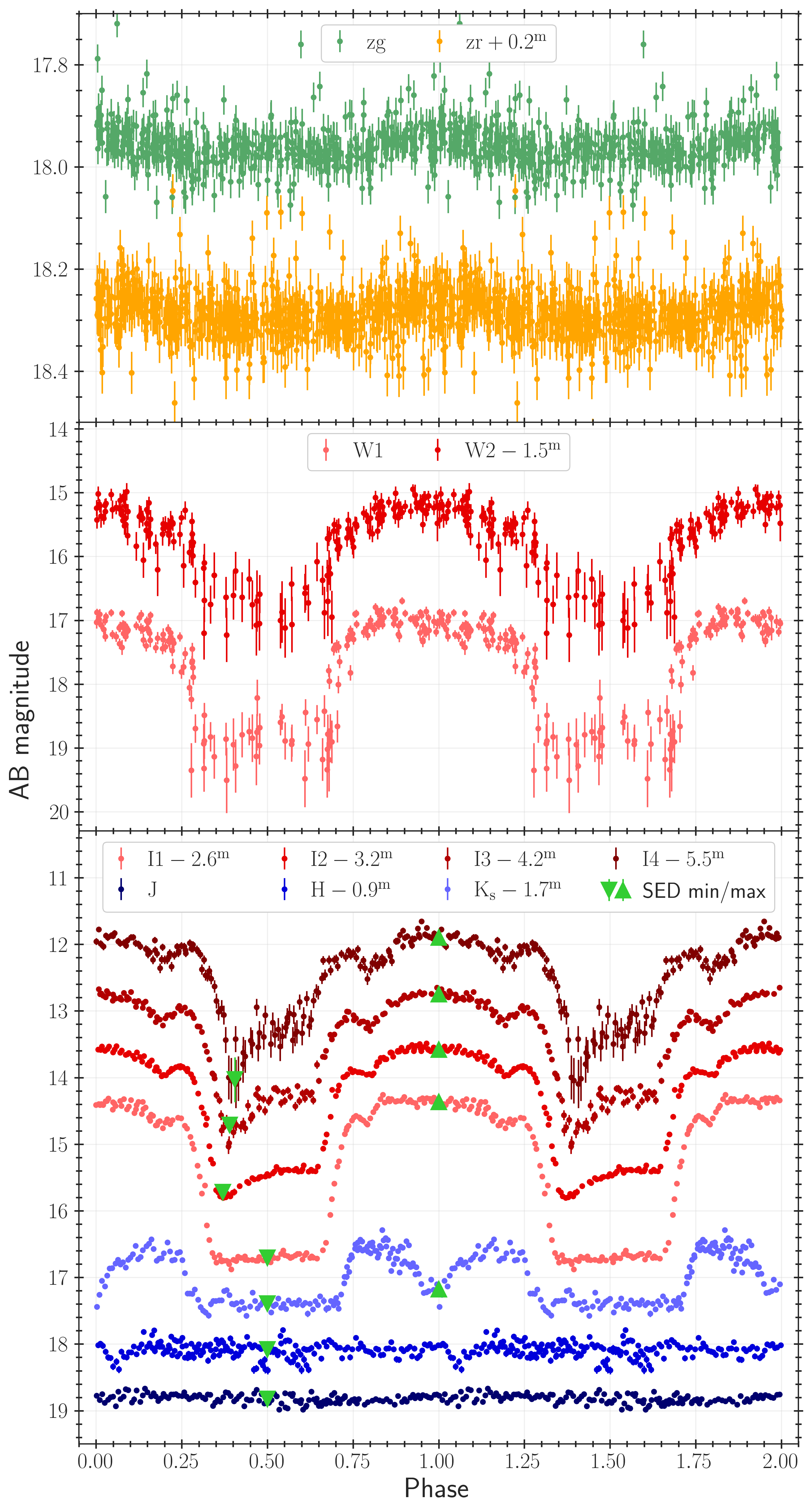}
    \vspace*{-7mm}
    \caption{Phase-folded optical and infrared light curves of {\obj}. Top panel: ZTF $g$- and $r$-band light curves. Middle panel: WISE W1 and W2 light curves from NEOWISE catalog, covering 9.44~years of observations. Bottom panel: infrared light curves in $J$, $H$, $K_s$ bands (ground-based) and IRAC $\mathrm{I1}$--$\mathrm{I4}$ bands from \textit{Spitzer}. Green triangles mark minimum fluxes and bright-phase midpoints ($\varphi=1$) used in the SED (Fig.~\ref{fig:sed_picture}).}
    \label{fig:phased_lc}
    \vspace*{3mm}
    \end{minipage}
\end{figure}

The component parameters of {\obj} were estimated through spectral energy distribution modeling. 
Ultraviolet fluxes were adopted from the GALEX light curves \citep{Linnell10} in the FUV ($\lambda_{\rm eff} \approx 1535$~\AA) and NUV ($\lambda_{\rm eff} \approx 2301$~\AA) bands, which show variability amplitudes of 19\% and 7\%, respectively. Additional ultraviolet data were obtained with the UVOT instrument on board the \textit{Swift} observatory. We performed aperture photometry on all available Swift/UVOT images using the \texttt{uvotmaghist} tool from the HEASoft\footnote{The HEASoft software package is available at https://heasarc.gsfc.nasa.gov/lheasoft/.} package, extracting fluxes in the UVW1 ($\lambda_{\rm eff} \approx 2628$\AA), UVM2 ($\lambda_{\rm eff} \approx 2249$\AA), and UVW2 ($\lambda_{\rm eff} \approx 2050$~\AA) filters. Optical and near-infrared fluxes were compiled from the SDSS catalog (bands $u$, $g$, $r$, $i$, $z$; \citealt{Ahumada20}) and the VISTA archive (bands $Z$, $Y$, $J$, $H$, $K_s$; \citealt{Edge13}). Since the SDSS and VISTA observations are not phase-resolved, they are affected by orbital modulation. According to \citet{Burleigh06}, the modulation amplitude decreases from 9.2\% in the $u$ band to 2.6\% in the $i$ band. To construct the near-infrared SED, we also included $J$, $H$, and $K_s$ photometry from \citet{Debes06}. In the mid-infrared, {\obj} was observed by the \textit{Spitzer} Space Telescope with the IRAC camera \citep{Harrison15}, providing calibrated images in the I1 ($3.6~\mu$m), I2 ($4.5~\mu$m), I3 ($5.8~\mu$m), and I4 ($8.0~\mu$m) bands and available via the IRSA\footnote{The Spitzer data archive is available at https://irsa.ipac.caltech.edu/\-data/\-SPITZER/\-docs/\-spitzerdataarchives/.} service. We extracted fluxes from the Spitzer/IRAC images using aperture photometry with the \texttt{photutils}\footnote{The photutils package is available at https://photutils.readthedocs.io/en/stable/.} package. The corresponding phase-folded light curves constructed using the ephemeris (\ref{ephemeris}) are shown in Fig.~\ref{fig:phased_lc}. The midpoint of the bright phase in the IRAC bands coincides well with those in the ZTF and WISE light curves.

\begin{figure*}[h!]
    \centering 
    \vspace*{2mm}
    \includegraphics[width=0.95\textwidth]{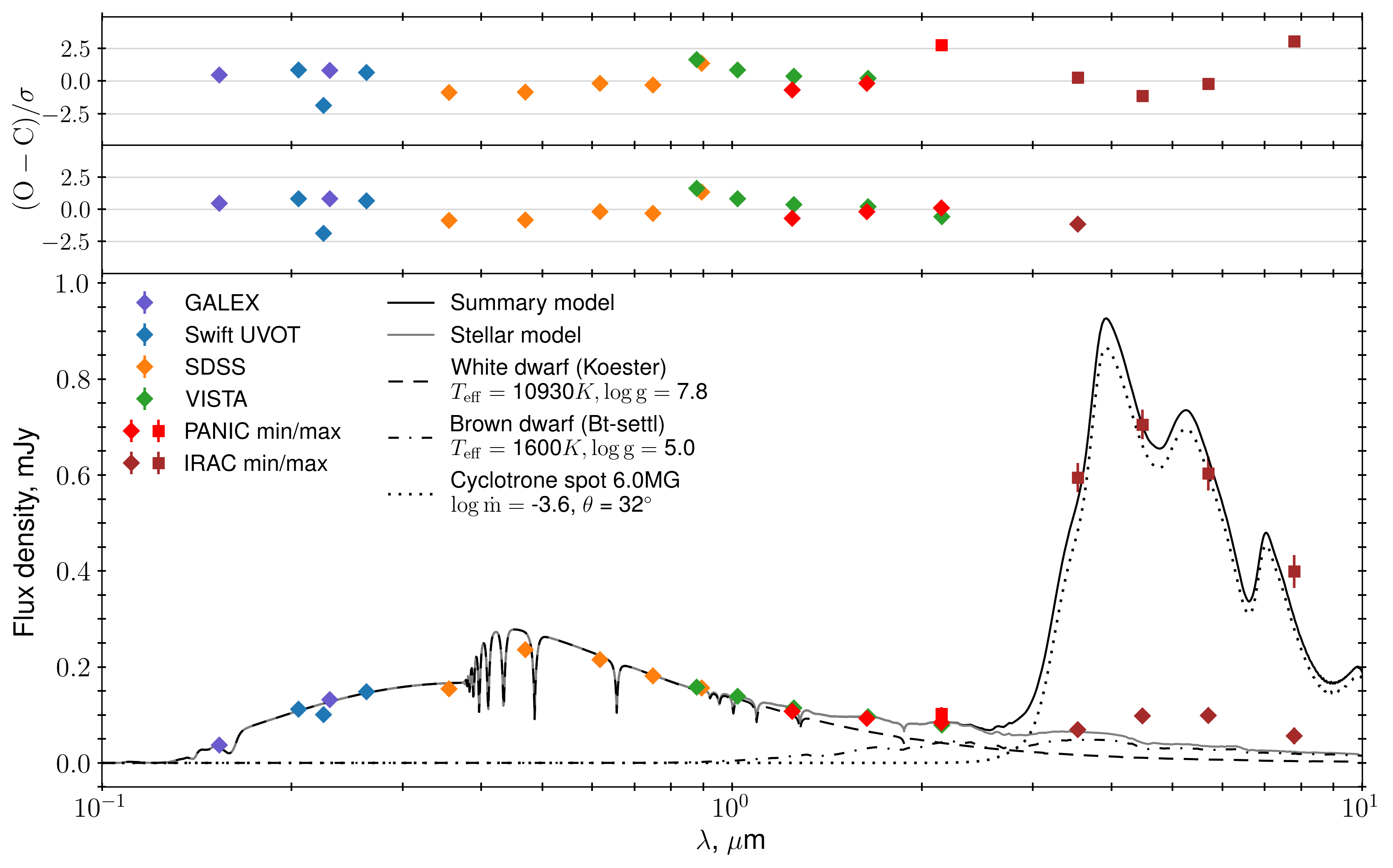} 
    \vspace*{-3mm}
    \caption{Bottom panel: observed SED of {\obj} (data points) compared to the best-fit model spectrum (solid line). The individual contributions from the white dwarf, the donor, and the cyclotron emission component are shown. Top panels: residuals $\mathrm{(O-C)/\sigma}$ between observed and model fluxes across the photometric bands. The lower subpanel displays residuals for the combined stellar spectrum (white dwarf + brown dwarf), while the upper subpanel additionally includes cyclotron emission from the accretion spot. Note that the SED in the bottom panel compares the model spectrum to monochromatic flux estimates, whereas the fitting was performed using band-integrated fluxes.}
    \label{fig:sed_picture}
\end{figure*}

The color excess along the line of sight to {\obj} was estimated using the 3D extinction maps from \citet{Leike20} and the \texttt{dustmaps}\footnote{The dustmaps library of interstellar extinction maps is available at https://dustmaps.readthedocs.io/en/latest/.} Python library. Assuming a distance of $d = 154 \pm 3$~pc based on the Gaia~DR3 parallax \citep{Vallenari22}, we obtained $E(B-V) = (9\pm8) \times 10^{-5}$. Despite the low extinction, all fluxes were corrected using the reddening law of \citet{Fitzpatrick99}. The resulting dereddened SED of {\obj} is shown in Fig.~\ref{fig:sed_picture}, and the fluxes used are listed in Table~\ref{table:fluxes}.

Starting from the $K_s$ band ($\lambda \gtrsim 2.2~\mu$m), {\obj} exhibits strong orbital variability caused by cyclotron emission from the accretion spot (see Fig.~\ref{fig:phased_lc}). To account for this, we considered both the faint and bright orbital phases when modeling the SED. The “faint phase” corresponds to orbital phase $\phi \approx 0.5$, when the accretion spot is hidden behind the white dwarf, while the “bright phase” ($\phi = 1$) corresponds to its full visibility. Infrared fluxes selected for each phase are indicated in Fig.~\ref{fig:phased_lc}. Notably, the duration of the bright phase increases with the passband wavelength $\lambda_{\rm eff}$, from $\phi \in [0.7, 1.3]$ in $K_s$ to $\phi \in [0.6, 1.4]$ in the I4 band. This behavior likely reflects the inhomogeneous structure of the accretion spot, with a hot inner core dominating the emission in the $K_s$ band and a cooler periphery contributing more significantly at longer wavelengths.

\begin{table*}
    \caption{Photometric data used to construct the SED of {\obj}. 
    The table provides 
    the instruments used for the observations, the effective wavelengths $\lambda_{\mathrm{eff}}$ 
    of the photometric filters, the observed fluxes $f_{\mathrm{obs}}$ corrected for interstellar extinction, and the flux uncertainties $\sigma$ adopted in the modeling. For the infrared filters $K_s$ and $\mathrm{I1}$--$\mathrm{I4}$, both the minimum fluxes measured during the faint phase (min) and the fluxes corresponding to the midpoint of the bright phase (max) are listed.}
    \label{table:fluxes}
    \begin{center}
    \begin{threeparttable}
        \begin{tabular}{lccc}
            \hline
            Instruments                & $\lambda_{\mathrm{eff}}$,~\AA  & $f_{\mathrm{obs}}$, $\mathrm{\mu Jy}$  & $\sigma$, $\mathrm{\mu Jy}$     \\ \hline
            GALEX $\mathrm{FUV}$           &  1535.1       & 36.9       & 7.0      \\  
            GALEX $\mathrm{NUV}$           &  2301.8       & 131.6      & 9.1      \\
            Swift/UVOT $\mathrm{UVW2}$     &  2049.9       & 111.8      & 11.3      \\          
            Swift/UVOT $\mathrm{UVM2}$     &  2248.6       & 101.0      & 11.2      \\
            Swift/UVOT $\mathrm{UVW1}$     &  2628.4       & 148.7      & 12.3      \\
            SDSS $u$              &  3556.5       & 154.2      & 14.5      \\
            SDSS $g$              &  4702.5       & 236.0      & 10.7      \\
            SDSS $r$              &  6175.6       & 215.2      & 8.3       \\
            SDSS $i$              &  7490.0       & 181.5      & 6.0       \\
            SDSS $z$              &  8946.7       & 155.9      & 5.6       \\
            VISTA $Z$             &  8793.6       & 157.9      & 3.4       \\
            VISTA $Y$             & 10210.7       & 138.7      & 6.3       \\
            VISTA $J$             & 12525.3       & 114.6      & 7.4       \\
            VISTA $H$             & 16432.8       & 96.6       & 10.7      \\
            VISTA $K_s$           & 21521.7       & 79.0       & 7.6       \\
            PANIC $J$\tnote{*}    & 12459.8       &	107.4      & 6.9       \\
            PANIC $H$\tnote{*}    & 16357.4       & 92.7       & 10.4      \\
            PANIC $K_s$ min/max\tnote{*} & 21503.0      & 83.8 / 102.3   & 7.0 / 13.9       \\
            Spitzer/IRAC $\mathrm{I1}$ min/max    & 35378.4      & 68.9 / 594.9   & 12.5 / 30.3     \\
            Spitzer/IRAC $\mathrm{I2}$ min/max    & 44780.5      & 98.0 / 705.4   & 2.2 / 30.7      \\
            Spitzer/IRAC $\mathrm{I3}$ min/max    & 56961.8      & 98.7 / 603.2   & 7.6 /  35.4     \\
            Spitzer/IRAC $\mathrm{I4}$ min/max    & 77978.4      & 55.9 / 399.2   & 17.4 / 34.2     \\
            \hline
        \end{tabular}
        \begin{tablenotes}
            \item[*] The fluxes were derived using data from \cite{Debes06}.
        \end{tablenotes}
    \end{threeparttable}
    \end{center}
\end{table*}

\section{Modeling}

The spectral energy distribution of {\obj} was modeled as the sum of the spectra from the stellar components and the cyclotron emission produced in the accretion spot. The white dwarf fluxes were calculated by interpolating a grid of theoretical spectra based on hydrogen, plane-parallel LTE atmospheric models \citep{Koester10}. The brown dwarf fluxes were derived by interpolating synthetic spectra from the BT-Settl models \citep{Allard12}. In both cases, the interpolation was performed over the effective temperature $T_{\mathrm{eff}}$ and the surface gravity $\log g$.

The cyclotron emission was modeled using the heated white dwarf atmosphere approximation, also known as the bombardment regime. This regime is applicable at low local accretion rates and in strong magnetic fields ($\dot{m} (B/10^7$G$)^{-2.6} \textless 0.1~\mathrm{g}~\mathrm{cm^{-2}}~\mathrm{s^{-1}}$, \citealt{Campbell08}), when the protons of the infalling gas lose their kinetic energy primarily through Coulomb interactions with atmospheric electrons. In such magnetic fields, electrons efficiently radiate the acquired energy via cyclotron emission, so that the proton stopping distance becomes comparable to their mean free path. As a result, no accretion shock forms, and the temperature of the accretion spot is strongly dependent on the local accretion rate.

We adopted the temperature profile across the shock region depth as proposed by \cite{Woelk93}, who parameterized it in terms of the local accretion rate $\dot{m}$, magnetic field strength $B$, and white dwarf mass $M_{\mathrm{wd}}$. This model was employed to solve the radiative transfer equation in the presence of strong Faraday rotation, which is characteristic of accretion spots in polars \citep{Rousseau96}. In this case, the polarized radiative transfer equations reduce to two independent equations for the ordinary ($o$) and extraordinary ($e$) polarization modes. The corresponding intensities were calculated under the LTE approximation as
\begin{gather*}
I_{o,e}(\omega, \theta) = \int^{x_s}_0 \frac{\alpha_{o,e}(\omega, \theta, T)}{2\rho \cos{\theta}} B(\omega, T) \times \\
\times \exp\bigg(-\int^x_0 \frac{\alpha_{o,e}(\omega, \theta, T)}{\rho \cos{\theta}}d\tilde{x}\bigg)dx ,
\label{transfer_solution}
\end{gather*}
where the integration is carried out over the column mass $x$, with $\rho = \rho(x)$ denoting the gas density, $T = T(x)$ the temperature profile, $B(\omega, T)$ the Planck function, $\alpha_{o,e}$ the cyclotron absorption coefficients for the two polarization modes, and $\theta$ the angle between the magnetic field direction and the line of sight. The cyclotron absorption coefficients $\alpha_{o,e}$ were calculated using the expressions provided by \cite{Chanmugam81}. The integration extends over the full depth of the accretion column, up to the column mass $x_s$. The total intensity is then given by $I = I_e + I_o$. A detailed description of the method used to compute the cyclotron emission spectrum and the temperature profile within this framework can be found in \citep{Rousseau96, Woelk93}.

The modeling of the spectral energy distribution was performed using the least-squares method, which involves minimizing the function
\begin{equation}
    \chi^2 = \displaystyle\sum_{k} \Big(\frac{f_{k} - f_{k}^{obs}}{\sigma_k}\Big)^2,
\label{sed_least_square}
\end{equation}
where the index $k$ runs over all photometric bands, $f_{k}^{\mathrm{obs}}$ is the observed flux corrected for interstellar extinction, $f_k$ is the corresponding theoretical flux, and $\sigma_k$ is the observational uncertainty. The ultraviolet and optical light curves of {\obj} exhibit rotational variability due to heating near the accretion spot on the white dwarf. Since this effect was not explicitly included in the observed flux calculations, the uncertainties $\sigma_k$ were taken to be equal to the semi-amplitude of the variations. The model flux $f_k$ was constructed as a combination of the fluxes from the white dwarf $f_{k}^{\mathrm{wd}}$, the brown dwarf $f_{k}^{\mathrm{bd}}$, and the cyclotron emission source $f_{k}^{\mathrm{cyc}}$, according to:
\begin{equation}
    f_{k} = \frac{1}{4} (\theta_{\mathrm{wd}}^2 f_{k}^{\mathrm{wd}} + \theta_{\mathrm{bd}}^2 f_{k}^{\mathrm{bd}} + \theta_{\mathrm{cyc}}^2 f_{k}^{\mathrm{cyc}}),
\label{sum_flux}
\end{equation}
where $\theta_{\mathrm{wd}}$, $\theta_{\mathrm{bd}}$, and $\theta_{\mathrm{cyc}}$ are the angular diameters of the white dwarf, the donor star, and the accretion spot, respectively. The theoretical fluxes $f_k$ were obtained by convolving the model spectra with the transmission profiles of the photometric filters. The $\chi^2$ minimization was performed using the Nelder--Mead simplex algorithm \citep{Gao12}.

The parameters of the stellar components and the accretion spot were derived using different subsets of observational data. To estimate the properties of the white dwarf and the donor, we used only those observations in which the contribution from cyclotron source is negligible. Cyclotron emission becomes insignificant at wavelengths $\lesssim 16000$~\AA, as indicated by the lack of noticeable variability in the $J$ and $H$ bands (see Fig.~\ref{fig:phased_lc}). The $K_s$ and I1 light curves exhibit plateau phases lasting $\approx {^1/_2} P_{\mathrm{orb}}$, during which the system reaches minimum flux. These phases likely correspond to the accretion spot being occulted behind the white dwarf’s limb, and the plateau fluxes can be used to estimate the parameters of the stellar components. Spitzer/IRAC observations in bands $I2$–$I4$ were not included due to the absence of a clearly defined plateau phase, possibly indicating the presence of a secondary accretion spot (see Fig.~\ref{fig:phased_lc}). The dominance of white dwarf emission in the ultraviolet and optical ranges allows for a reliable determination of its atmospheric parameters: an effective temperature $T_{\mathrm{eff}} = 10930 \pm 350$~K and a surface gravity of $\log g = 7.8 \pm 0.4$. It should be noted that $\log~g$ was determined solely from the shape of SED, while the angular diameter $\theta_{\mathrm{wd}}$ was treated as an independent fitting parameter. The derived temperature is consistent with the earlier estimate of $T_{\mathrm{eff}} = 10000 \pm 1000$~K reported by \cite{Schmidt05a}. Assuming a fixed surface gravity of $\log g = 5.0$, the effective temperature of the secondary  was estimated to be $T_{\mathrm{eff}} = 1600 \pm 180$~K, consistent with a brown dwarf of spectral type L6 --- L8. This result is consistent with the findings of \cite{Farihi08}, who inferred an L8-type donor based on infrared spectroscopy. A comparison between the observed and modeled SED is shown in Fig.~\ref{fig:sed_picture}. At brightness minimum, the infrared fluxes in the $I2$ -- $I4$ bands exhibit an excess relative to the combined emission from the stellar components, which may suggest the presence of a secondary cyclotron-emitting region.

Using the parallax of {\obj} and the previously determined angular diameters, we derive radii of $R_1 = 0.012 \pm 0.001~R_{\odot}$ and $R_2 = 0.095 \pm 0.018~R_{\odot}$ for the white dwarf and brown dwarf, respectively. Applying the mass–radius relation from \cite{Nauenberg72}, we estimate a white dwarf mass of $M_1 = 0.61 \pm 0.05~M_{\odot}$. The corresponding surface gravity $\log g = 8.0 \pm 0.2$, is in good agreement with the value inferred from the SED. The donor’s radius also agrees with the estimate $R_2 \approx 0.09~R_{\odot}$ reported by \cite{Farihi08}. The derived parameters of the white dwarf and the donor are summarized in Table~\ref{table:pars}.

\begin{table*}
    \caption{Physical parameters of the white dwarf and donor in {\obj}.}
    \label{table:pars}
    \begin{center}
    \begin{threeparttable}
        \begin{tabular}{lcc}
            \hline
            Parameters                & White dwarf         & Donor  \\ \hline
            $T_{\mathrm{eff}}$ [K]    & $10930 \pm 350$     & $1600 \pm 180$     \\
            $\log (\mathrm{g [cm/s^2]}$)                 & $8.0 \pm 0.2$     &                    \\  
            $\theta''$            & $(3.7 \pm 0.2) \times 10^{-7}$  & $(2.9 \pm 0.6) \times 10^{-6}$ \\
            $R/R_{\odot}$           & $0.012 \pm 0.001$ & $0.095 \pm 0.018$   \\
            $M/M_{\odot}$         & $0.61 \pm 0.05$   & $0.014 - 0.065$\tnote{*}           \\
            \hline
        \end{tabular}
        \begin{tablenotes}
            \item[*] Donor mass range assumes Roche-lobe filling.
        \end{tablenotes}
    \end{threeparttable}
    \end{center}
\end{table*}

To estimate the parameters of the accretion spot, we used infrared fluxes measured at the midpoint of the bright phase ($\varphi = 1$). The fitting was performed with fixed white dwarf and donor parameters. The best-fit model corresponds to a local accretion rate of $\log \dot{m} = -3.6 \pm 0.2$ ($\dot{m}$ in units of $\mathrm{g}~\mathrm{cm^{-2}}~\mathrm{s^{-1}}$), a viewing angle of $\theta = 32^{\circ} \pm 6^{\circ}$, and a magnetic field strength of $B = 6.0$~MG. The discrepancy with the $B \approx7$~MG estimate by \cite{Farihi08} may arise from differences in cyclotron emission models or spectral coverage. A similar magnetic field of $\approx 7$~MG was also inferred by \cite{Schmidt05a} from Zeeman splitting, which varies with the white dwarf’s rotational phase. We note that varying $B$ within the $6-7$~MG range has only a minor effect on the inferred accretion rate ($\Delta \log \dot{m} < 0.2$). A comparison of the observed and modeled fluxes at the midpoint of the bright phase is shown in Fig.~\ref{fig:sed_picture}.

The total accretion rate can be estimated from the angular diameter of the cyclotron source and the local accretion rate using $\dot{M} = \dot{m} S_{\mathrm{spot}}$, where $S_{\mathrm{spot}} \approx \pi \theta_{\mathrm{cyc}}^2 d^2 / 4$ is the area of the accretion spot. This yields an accretion rate of $\dot{M} = (6.4 \pm 2.8) \times 10^{-13}~M_{\odot}/\mathrm{yr}$.
Alternatively, the accretion rate can be derived from the accretion luminosity:
\begin{equation}
L_{\mathrm{acc}} = \frac{G M_1 \dot{M}}{R_1},
\label{accretion_luminosity}
\end{equation}
where $G$ is the gravitational constant. An unabsorbed X-ray flux of $F_{\mathrm{x}} = 1.5 \times 10^{-13}$~erg~cm$^{-2}$~s$^{-1}$ was measured by \cite{Stelzer17}, consistent with the value of $F_{\mathrm{x}} \approx 1.2 \times 10^{-13}$~erg~cm$^{-2}$~s$^{-1}$ obtained by \cite{Burleigh06} using Swift/XRT. Additionally, {\obj} was detected in the eROSITA all-sky survey aboard \textit{Spektr-RG} \citep{Munoz23}, with $F_{\mathrm{x}} \approx 0.75 \times 10^{-13}$~erg~cm$^{-2}$~s$^{-1}$. Since the value reported by \cite{Stelzer17} is phase-averaged, the actual flux from the accretion spot (based on the X-ray light curve) is estimated to be $F_{\mathrm{x}} = 3 \times 10^{-13}$~erg~cm$^{-2}$~s$^{-1}$. The modeled cyclotron flux is $F_{\mathrm{cyc}} = 3.5 \times 10^{-13}$~erg~cm$^{-2}$~s$^{-1}$. Together, they yield a total accretion luminosity of $L_{\mathrm{acc}} = L_{\mathrm{x}} + L_{\mathrm{cyc}} = 1.84 \times 10^{30}$~erg/s, corresponding to an accretion rate of $\dot{M} = (3.1 \pm 0.3) \times 10^{-13}~M_{\odot}/\mathrm{yr}$ via Eq.~(\ref{accretion_luminosity}). In both methods, the inferred accretion rate is significantly higher than the earlier estimate of $3.4 \times 10^{-14}~M_{\odot}/\mathrm{yr}$ reported by \cite{Stelzer17}.
In polars, the fractional area of the accretion spot relative to the white dwarf surface, $f = S_{\mathrm{spot}} / 4\pi R_{\mathrm{wd}}^2$, typically lies in the range $10^{-5} - 10^{-3}$. For {\obj}, we obtain $f = 0.017 \pm 0.008$ based on the size of the cyclotron source.

\section{Discussion}

\begin{figure}[h!]
    \centering
    \begin{minipage}[t]{\columnwidth}
      \centering
      \includegraphics[width=1.0\linewidth]{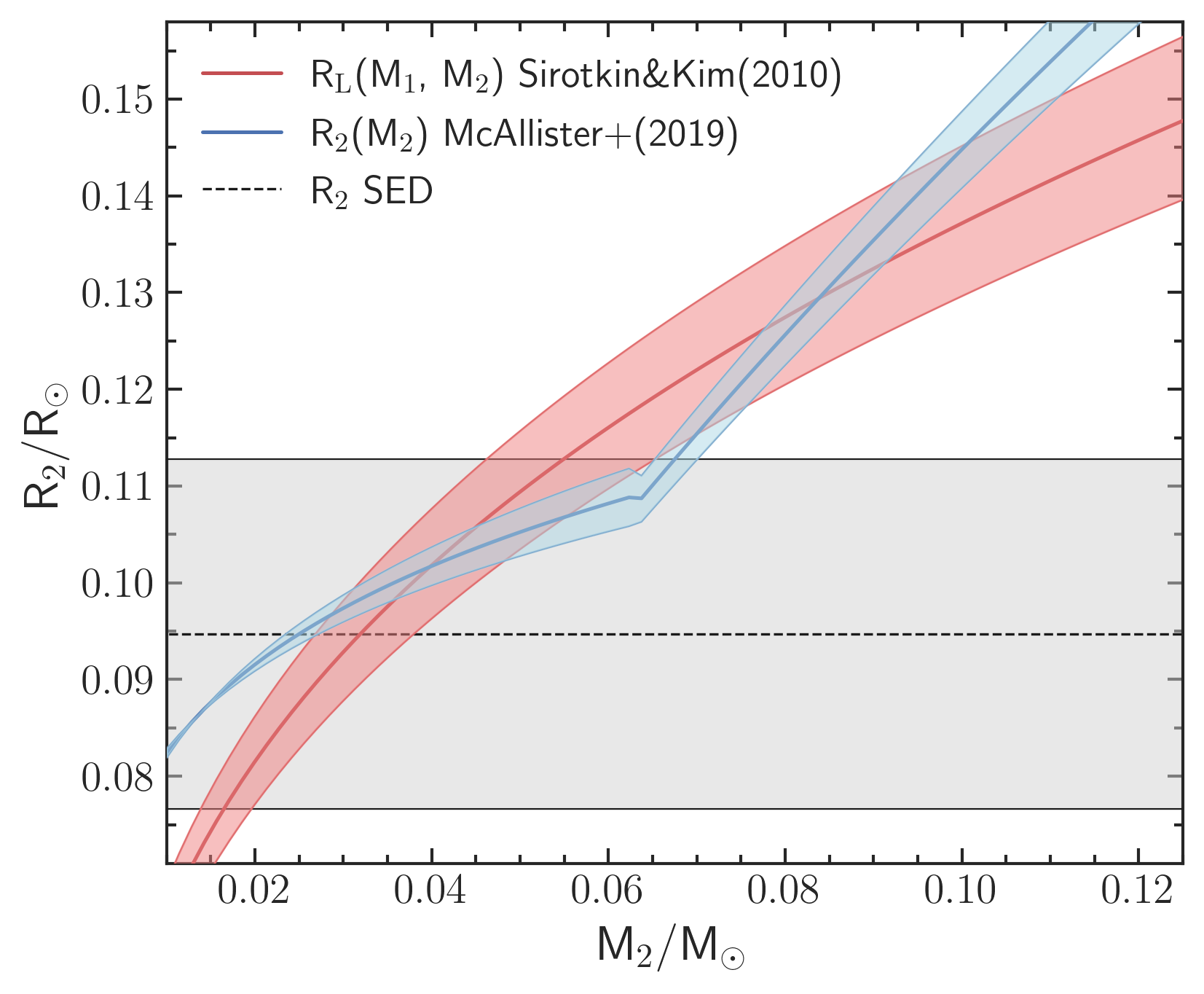}
      \vspace*{-5mm}
      \caption{The effective Roche lobe radius $R_L$ as a function of donor mass $M_2$ (red line). The gray shaded area represents the range of donor radii $R_2$ inferred in this study. The blue line corresponds to the semi-empirical $R_2-M_2$ relation from \cite{McAllister19}.}
      \label{fig:rad2_mass2}
    \end{minipage}
\end{figure}

According to \cite{Sirotkin10}, the effective radius $R_L$ of a star filling its Roche lobe can be estimated by the relation
\begin{equation}
R_L=A\frac{0.5126 q^{0.7388}}{0.6710q^{0.7349}+\mathrm{ln}(1+q^{0.3983})},
\label{eq_rl}
\end{equation}
where $A=(M_2(1+1/q)P_{\mathrm{orb}}^2)^{1/3}$ is the semi-major axis of the binary, and $q=M_2/M_1$ is the mass ratio. This expression refines the classical approximation by \cite{Eggleton83} for polytropic stellar models with an index $n=3/2$, which better represent fully convective donor. Fig.~\ref{fig:rad2_mass2} shows the Roche lobe radius of the donor as a function of its mass, calculated assuming a white dwarf mass of $M_1 = 0.61 \pm 0.05$~M$_{\odot}$ for {\obj}. It is evident that the donor mass must exceed 0.014~M$_{\odot}$. By considering the upper mass limit of 0.065~M$_{\odot}$ for period bouncers \citep{McAllister19}, we obtain a lower limit on the Roche lobe filling factor $R_2/R_L \gtrsim 0.67$.


It is interesting to compare the accretion rate we derived for {\obj} with the expected evolutionary accretion rate driven by angular momentum loss through gravitational wave radiation. According to \cite{Webbink05}, the accretion rate from a donor filling its Roche lobe has a lower limit given by
\begin{equation}
    | \dot{M} | \ge \frac{M_2}{\tau_{gr}} \Big( \frac{2}{3} - \frac{M_2}{M_1} \Big)^{-1},
\label{evol}
\end{equation}
where the timescale for angular momentum loss due to gravitational wave emission $\tau_{gr}$, is expressed as
\begin{equation}
    \tau_{gr}^{-1} = \frac{32}{5} \frac{G^{5/3}}{c^5} \frac{M_1 M_2}{(M_1+M_2)^{1/3}} \Big( \frac{2\pi}{P_{\mathrm{orb}}} \Big)^{8/3}.
\label{tau_grav}
\end{equation}
Substituting the parameters derived for {\obj} into equations (\ref{evol}) and (\ref{tau_grav}) yields a lower limit on the accretion rate of $\dot{M} \ge 2\times 10^{-11}$~M$_{\odot}$/yr. This value exceeds our measured accretion rate of $\dot{M} \approx 3\times 10^{-13}$~M$_{\odot}$/yr by nearly two orders of magnitude.

Two possible explanations may account for the low accretion rate observed in {\obj}. The first is that {\obj} is currently in a low state. It is well established that long-term light curves of polars exhibit irregular transitions between high and low states, typically differing by several magnitudes. These transitions are associated with changes in the mass accretion rate onto the white dwarf. Although the underlying cause of these changes remains debated, the most widely accepted explanation involves suppression of mass transfer from the donor due to its magnetic activity \citep{Livio94, King98}. In particular, when starspots accumulate near the inner Lagrange point L$_1$, their magnetic fields may inhibit mass loss, leading the system into a low state.

The long-term light curve of {\obj} shows no significant brightness variations over the past $\approx 20$ years of observations. If the low accretion rate in {\obj} does indeed correspond to a low state, it would be unusually long-lasting compared to most known polars. However, there are known cases of polars remaining in low states for extended periods. For example, the short-period ($P_{\mathrm{orb}} \approx 87$~min) polar EF~Eri remained in a low state for at least 10~years \citep{Schwope07}. Similarly, the polars IL~Leo ($P_{\mathrm{orb}} = 82$~min) and SDSS~J0837+3830 ($P_{\mathrm{orb}} = 191$~min), previously classified as LARP systems, were recently observed by ZTF to transition into high states \citep{Roestel24}. It is also worth noting that the accretion rate we derived is consistent with those measured in polars during low states. For instance, in the low state of AR~UMa, \cite{Gansicke01} estimated an accretion rate of $\dot{M} \sim 1.7 \times 10^{-13}~M_{\odot}$/yr. \cite{Schwope07} reported an accretion luminosity for EF~Eri in the low state of $L_{\mathrm{acc}} = 2.4 \times 10^{30}~(d/100,\mathrm{pc})^2$~erg/s, corresponding to $\dot{M} \sim 5 \times 10^{-13}M_{\odot}$/yr for a distance of $d = 160$~pc and a white dwarf mass of $M_1 \sim 0.9~M_{\odot}$ \citep{Schwope10}. \cite{Ramsay04} estimated the X-ray luminosity of a sample of polars in low states to be $L_X \sim 10^{30}$~erg~s$^{-1}$, in agreement with the value observed for {\obj}.

According to \cite{Webbink05}, stellar winds in L--M dwarfs are driven by atmospheric heating from coronal X-ray emission. \cite{Stelzer17} placed an upper limit on the X-ray luminosity of {\obj} during brightness minimum (when the accretion spot is eclipsed by the white dwarf), finding $L_X \lesssim 10^{27}$~erg~s$^{-1}$, which can be adopted as a constraint on coronal emission. Based on the relations in \cite{Webbink05}, this implies a stellar wind mass-loss rate of $\sim 10^{-15}~M_{\odot}$/yr, about two orders of magnitude lower than the accretion rate we have estimated. Determining the mass-loss rate from brown dwarf winds remains challenging. Recently, \cite{Walters23} analyzed spectra of white dwarfs whose atmospheres are “polluted” by wind from their brown dwarf companions. They constrained the wind-driven mass-loss rate of L-type dwarfs of $\dot{M} \lesssim 5 \times 10^{-17}~M_{\odot}$/yr. These results support the interpretation that {\obj} is currently in an extended low-accretion state, rather than accreting via a stellar wind. If the donor in {\obj} fills its Roche lobe, its mass must lie in the range $M_2 \in 0.014 - 0.065~ M_{\odot}$ (see Fig.~\ref{fig:rad2_mass2}). Assuming the semi-empirical $R_2-M_2$ relation for magnetic systems proposed by \cite{McAllister19} holds, the donor in {\obj} is likely has a mass of $M_2 \approx 0.04~M_{\odot}$ and an effective radius of $R_2 \approx 0.1~R_{\odot}$ (see Fig.~\ref{fig:rad2_mass2}).

In our view, modeling the spectral energy distribution in LARP systems offers a promising approach. Indeed, in the case of {\obj}, SED modeling yielded an accretion rate consistent with the value inferred from accretion luminosity (i.e., the combined X-ray and cyclotron emission). This analysis required infrared data from the \textit{Spitzer} Space Telescope. However, for systems with stronger magnetic fields ($\sim 30 - 70$~MG), a significant fraction of the cyclotron emission may fall within the optical and near-infrared range accessible to ground-based observations. Additionally, the intrinsically low X-ray luminosity of LARP systems poses challenges for their X-ray monitoring.

\section{Acknowledgements}

This study was supported by the Russian Science Foundation, grant No. 22-72-10064 (https://rscf.ru/project/22-72-10064/).

\end{document}